\documentclass[10pt]{article}

\usepackage{html}
\usepackage{hthtml}
\usepackage{footnpag}

\begin{document}

\title{
\begin{flushright}
\normalsize{FTUV-03-0902} 
\vspace{0.5cm}
\end{flushright}
JaxoDraw: \\ A graphical user interface for drawing Feynman diagrams}

\author{D. Binosi and L. Theu{\ss}l \\ 
Departamento de F\'{\i}sica Te\'orica, Universidad de Valencia, \\
E-46100 Burjassot (Valencia), Spain}
\date{September 1, 2003}

\maketitle

\begin{abstract}

JaxoDraw is a Feynman graph plotting tool written in Java. 
It has a complete graphical user interface that allows all actions to 
be carried out via mouse click-and-drag operations in a WYSIWYG fashion. 
Graphs may be exported to postscript/EPS format and can be saved in XML 
files to be used in later sessions. One of the main features of JaxoDraw is the
possibility to produce \LaTeX~code that may be used to generate graphics 
output, thus
combining the powers of \TeX/\LaTeX~with those of a modern day drawing program.
With JaxoDraw it becomes possible to draw even complicated Feynman diagrams 
with just a few mouse clicks, without the knowledge of any programming
language. 

\end{abstract}

\pagebreak

\section{Introduction}
\label{intro}

It is a widely accepted convention today in the scientific community to write
scientific papers using the \TeX~and \LaTeX~environments. The high quality,
publication-style typesetting of \LaTeX~has made it now a de facto standard, to
such an extent that some scientific journals only accept submission of papers 
in electronic form anymore. The portability goal of \TeX~ has however the
drawback that graphical representations are only possible in very rudimentary
form (using the \LaTeX~{\tt picture} environment or packages using a similar
approach). Useful as they are, mostly they are
too simple to draw complicated Feynman diagrams as needed in wide parts of
theoretical nuclear and particle physics today. 

This problem has led to the development of more sophisticated programs in the
past. The \htlink{UK List of \TeX Frequently Asked Questions}
{http://www.tex.ac.uk/cgi-bin/texfaq2html?label=drawFeyn} lists four
possibilities to draw Feynman diagrams in conjunction with \LaTeX:
Michael Levine's {\tt feynman}~\cite{Levine:1990} bundle; 
Jos Vermaseren's {\tt axodraw}~\cite{Vermaseren:1994je} package which uses 
PostScript specials and is thus slightly less portable but much more powerful;
Thorsten Ohl's {\tt feynmf}~\cite{Ohl:1995kr} package for \LaTeX2e which uses 
{\tt METAFONT} (or {\tt MetaPost}) to combine flexibility and portability;
and Norman Gray's 
\htlink{{\tt feyn}}{http://www.astro.gla.ac.uk/users/norman/distrib/latex/} 
package. These are all available from the \htlink{CTAN}{http://www.ctan.org/} 
archives.

Powerful as they are, all these methods have the common drawback that they
require some 'hard-coding' from the user side, there does not exist any graphical
user interface while modern day drawing programs 
do not include special options 
that are necessary to draw Feynman diagrams with the same quality as the one
achieved by \TeX/\LaTeX.

Our program JaxoDraw is an attempt to close that gap. As the name suggests, it
was initially meant to be a graphical user interface for Jos Vermaseren's 
{\tt axodraw} package, but it may be used independently of it. However, it is in
conjunction with {\tt axodraw} that JaxoDraw develops its main capabilities 
because of the possibility to combine the powers of \TeX/\LaTeX~with a modern 
drawing program. The main design goal of JaxoDraw was convenience and
ease-of-use; it should be possible for anybody to draw even complicated Feynman 
diagrams with just a few mouse clicks, without the knowledge of any programming 
language. JaxoDraw was written in Java, which means that (in principle)
it can be used on any platform where a Java Runtime Environment is installed.
This makes it completely portable, even though in the present version, 
the program includes the execution of some external commands that are inherently
system dependent and currently only work under certain operating systems.

This paper attempts to give a complete overview of JaxoDraw from installation
and usage instructions to documentation issues and possible developments.

\subsection{Overview}
\label{Overview}

JaxoDraw is a Feynman graph plotting tool written in Java. 
It has a complete graphical user interface that allows all actions to 
be carried out via mouse point-and-click-and-drag operations. Graphs may be 
exported to postscript / EPS format and can be saved in XML files to be used 
for later sessions. One of JaxoDraw's main features is the possibility to 
create \LaTeX~files which make use of J. Vermaseren's {\tt axodraw} package to create 
a Feynman diagram via {\tt latex} and {\tt dvips}. It was in fact the main motivation to 
write JaxoDraw, to create a graphical user interface for the {\tt axodraw} package.

A few of JaxoDraw's main features are:  

\begin{itemize}
\item[] Complete point-and-click graphical user interface. 
\item[] Pre-defined line styles for common particle representations. 
\item[] Saving and reading of sessions in XML format.
\item[] Export to Postscript, EPS and \LaTeX format.
\item[] Setting of permanent preferences.
\end{itemize}

\noindent
The home of JaxoDraw sources and documentation is:

\verb+http://altair.ific.uv.es/~JaxoDraw/home.html+,

\noindent
please refer to this page for (hopefully) up to date
information. \\
There is also a direct link to the JaxoDraw home page in
the web page of the Physics Department of the University of Valencia:

\htmlurl{http://fisteo.uv.es/catala/deptlinks.html}


\subsection{Installation}
\label{Installation}

You may download JaxoDraw in three different forms ({\tt xxx} denotes the
version number and platform type):

\medskip
\begin{tabular}{lcl}
{\tt JaxoDraw-xxx-src.tar.gz} &:& \parbox[t]{27em}{A gzipped tar file ($\sim$0.4 Mb) 
                                 containing the JaxoDraw
                                  sources} \\
{\tt JaxoDraw-xxx-bin.tar.gz} &:& \parbox[t]{27em}{A gzipped tar file ($\sim$0.6 Mb) 
                                  containing  precompiled binaries} \\
{\tt JaxoDraw-xxx-jre.tar.gz} &:& \parbox[t]{27em}{A gzipped tar file ($\sim$22 Mb) 
                                  containing precompiled binaries and a 
                                  Java Runtime Environment}
\end{tabular}

\medskip
\noindent
If you have a Java Developer Kit installed on your system and
you want to compile JaxoDraw yourself from sources, you may download the
{\tt src.tar.gz} file above. Check the prerequisites section~\ref{Prerequisites} and the
compiling from sources section~\ref{Compilesources} below.

\noindent
If you have a Java Runtime Environment installed on your system (or a Developer
Kit which includes the Runtime Environment), you may 
download the {\tt bin.tar.gz} file. Check the prerequisites 
section~\ref{Prerequisites} 
and section~\ref{Running} for information on how to start the program. 

\noindent
If you don't have Java installed or you are not sure about it, you should get 
the {\tt jre.tar.gz} file. This one contains a complete Java Runtime Environment 
({\tt jre}) so you won't need anything else anymore 
to run JaxoDraw. Note however the larger size of this last file.
Check section~\ref{Running} for information on how to start the program.

\subsubsection{Prerequisites}
\label{Prerequisites}
{\bf Note:} The following paragraph about the Java environment 
only applies if you downloaded either the source 
code {\tt src.tar.gz} or precompiled binaries {\tt bin.tar.gz}. The {\tt jre} 
version of JaxoDraw includes a Java Runtime Environment, just make sure then
that you have enough disc space and memory to run Java programs (a minimum of 
128Mb RAM is recommended).

\medskip
\noindent
Compilation and execution of JaxoDraw requires an installed and configured Java
environment on your system. 
To execute JaxoDraw you need a Java Runtime Environment ({\tt jre}), while for
compilation you need the Java Developer Kit ({\tt jdk}, which includes the 
{\tt jre}). The program was written with the SUN J2SDK
developer kit, version {\tt 1.4.1\_01}, using the SUN {\tt javac} compiler. 
We do not guarantee 
that the program compiles or runs with any earlier version or any other compiler 
(but we would like to hear from people who have succeeded to compile and run it
on different systems). 
The Java Developer Kit and Runtime Environment are free software and may be 
obtained from \htlink{SUN's web pages}{http://java.sun.com}. Please refer to 
SUN's Java pages for information on how to install and configure the Java 
environment on your system.

In order to profit from the \LaTeX~export file format, you need (apart from a
working \LaTeX~distribution, we use TeTeX 1.0.7) J. Vermaseren's 
\htlink{{\tt axodraw}}{http://www.nikhef.nl/~form/FORMdistribution/axodraw/}
package. This is now distributed along with JaxoDraw (with kind permission of
the author) but you have to install {\tt axodraw} on your system independently 
of JaxoDraw as described in appendix~\ref{axodraw}. Please refer to the 
{\tt axodraw} user guide for documentation on the package.

{\bf Note}: If you have already an {\tt axodraw.sty} installed on your system,
make sure to have a recent version (as the one included in the
downloads). Earlier versions did not
include some color macros that are necessary for the use with JaxoDraw (see
the Bugs section~\ref{Bugs}).

If you want to use the postscript preview option of JaxoDraw, you need to
specify an external postscript viewer. 

\subsubsection{Unpacking the archives}
\label{Unpacking}
Any of the packages available for download are unpacked with the command

{\tt tar -zxf JaxoDraw-xxx-zzz.tar.gz }

\noindent
under Linux or with the {\tt unzip} utility under Windows.
Here {\tt xxx} is the version number and {\tt zzz} is either {\tt src}, 
{\tt bin} or {\tt jre}. This will create a directory named 
{\tt JaxoDraw-xxx} (the JaxoDraw home directory) in the current directory. 
For the {\tt src} distribution the JaxoDraw home directory has the following 
structure:

\medskip
\begin{tabular}{ll}
{\tt src/}            & Source Files \\
{\tt JaxoDraw/}       & Binaries\\
{\tt JaxoDraw/icons/} & Icons\\
{\tt JaxoDraw/doc/}   & Documentation (User guide, man pages)\\
{\tt javadoc/}        & API Specification
\end{tabular}

\medskip
\noindent
where the {\tt JaxoDraw/} and  {\tt javadoc/} directories contain no 
regular files in the beginning. There are a number of files, like {\tt README}, 
{\tt BUGS}, etc., and some bash shell scripts that may be used to compile and 
run the program (see below).

The binary {\tt bin}  distribution does not contain any sub-directories but
includes an executable {\tt .jar} file in the JaxoDraw home directory. 
Also, with respect to the {\tt src} distribution there are a few files missing
that are only required for compilation/documentation issues.

The {\tt jre}  distribution has the same structure as the {\tt bin} distribution
but there is an additional directory 
{\tt j2re1.4.1\_03} (under Linux) or an executable 
{\tt j2re-1\_4\_1\_03-windows-i586.exe} (under Windows) containing the 
Java Runtime Environment.

\subsubsection{Compiling the sources}
\label{Compilesources}
If you have downloaded the source distribution of JaxoDraw, the {\tt JaxoDraw/} 
and  {\tt javadoc/} directories in the JaxoDraw home directory will contain no 
regular files in the beginning. Under Linux, you compile the sources with

{\tt javac -d . src/*.java }

\noindent
This creates the binary {\tt .class} files in the {\tt JaxoDraw/} sub-directory.
In order to create an executable {\tt .jar} file, issue the command

{\tt jar -cmf mainclass JaxoDraw.jar JaxoDraw/}

\noindent
This creates an executable {\tt JaxoDraw.jar} file in the JaxoDraw home 
directory.

\noindent
Under Windows, the procedure is basically the same as above, just replace the
Unix file separator {\tt /} by the one of Windows: {\tt $\backslash$}, and you
may have to give the absolute path to the {\tt javac} and {\tt jar} executables. 

\subsubsection{Creating the javadoc API specification}
\label{CreateAPI}
{\bf Note:} This is only needed if you are interested in the structure of the 
JaxoDraw source code, it is not required in order to run the program.

\medskip
\noindent
To create javadoc:

{\tt javadoc -d javadoc/ [-link Java-API] -link . src/*.java}

\noindent
where the optional {\tt Java-API} could be 
{\tt http://java.sun.com/j2se/1.4/docs/api}
to link against the online Java documentation from SUN, 
or a local directory if the javadoc is installed locally, for instance:

{\tt javadoc -d javadoc/ -link /usr/java/j2sdk1.4.1\_01/docs/api/ -link . src/*.java}

\medskip
\noindent
In any case, the command has to be run twice in order to get the 
cross-references right. This will create the javadoc API specification in the
{\tt javadoc/} sub-directory.

\subsubsection{Running the program}
\label{Running}
If you compiled the package yourself from sources 
(see sect.~\ref{Compilesources} above), you can
start JaxoDraw with the command line

{\tt java JaxoDraw/JaxoDraw}

\noindent
in the distribution home directory, or by

{\tt java -jar JaxoDraw.jar}

\noindent
if you want to use the binary {\tt .jar} file.

\noindent
If you downloaded the precompiled binary distribution
({\tt .bin}), just type

{\tt java -jar JaxoDraw.jar}

\noindent
or under Windows simply double click on the {\tt jar} file icon.

\noindent
If you downloaded the {\tt .jre} distribution and you are using Linux, 
you have to give the path to the
java executable in the j2re directory, typically:

{\tt ./j2re1.4.1\_03/bin/java -jar JaxoDraw.jar}.

\noindent
Under Windows you simply double click on the 
{\tt j2re-1\_4\_1\_03-windows-i586.exe} file to install the Java Runtime
Environment, followed by a double click on the {\tt jar} file icon to start
JaxoDraw.

\subsubsection{Using the shell scripts under Linux}
\label{Usingshellscripts}
Alternatively to the above procedure, under Linux you may use the bash shell 
scripts in the JaxoDraw home directory to compile and run the program 
(provided you have bash and Java installed on your system):

\medskip
\noindent
To compile the sources and create the binary classes:
  
  {\tt ./compile.bsh}

\noindent
To create the API specification:
  
  {\tt ./doc.bsh [-link Java-API]}

\noindent
where {\tt Java-API} is the same as in section \ref{CreateAPI}.

\noindent
Use the clean.bsh script to delete all binaries and javadoc files:
  
  {\tt ./clean.bsh}

\noindent
You may give a {\tt -all} option to this script which will also remove the 
{\tt JaxoDraw.jar} file (this may be recreated with the {\tt ./compile.bsh}
script).

\medskip
Note that {\tt compile.bsh}, {\tt doc.bsh } and {\tt clean.bsh} are only
included in the {\tt src} distribution.

\medskip
\noindent
To run the program:

  {\tt ./runme.bsh}

\noindent
or simply
  
  {\tt ./jaxodraw}

\noindent 
(which is a symlink to {\tt ./runme.bsh}).

\subsubsection{Installing JaxoDraw system wide}
\label{systemwide}

{\bf Note}: You need root privileges to install JaxoDraw system wide.

\noindent 
Execute the install shell script

  {\tt ./install.bsh}

\noindent 
This will put a {\tt JaxoDraw-xxx} directory into {\tt /usr/local}, a symlink 
to the executable bash script into {\tt /usr/local/bin} and a man page into
{\tt /usr/local/man/man1} (you may specify different locations by editing the 
{\tt ./install.bsh} file). Provided, {\tt /usr/local/bin} is in your {\tt PATH},
any user will then be able to run JaxoDraw just by typing {\tt jaxodraw}.
To uninstall, just use the script

  {\tt ./uninstall.bsh}


\subsection{Comments and bug reports}
\label{Comments}

Please send your comments, questions or bug reports to any of our e-mail 
addresses:
\begin{itemize}
\item[] \htmailto{Daniele.Binosi@uv.es}
\item[] \htmailto{Lukas.Theussl@uv.es}
\end{itemize}
When reporting bugs, you should be as specific as possible about the problem so
that we can easily reproduce it. Include some information about your operating
system and the version of JaxoDraw that you are using. Include for instance the
output of the 

{\tt java JaxoDraw/JaxoDraw \verb~--~info}  

\noindent
and 

{\tt java JaxoDraw/JaxoDraw \verb~--~version} 

\noindent
commands (this information is also available under the Help menu of the 
graphical user interface). If you are having problems with the
\LaTeX~compilation process, also include detailed information about your
\LaTeX~distribution, the version of {\tt dvips}, your postscript viewer and any
other information that may be relevant. Note in particular that there seem to
several versions of the {\tt axodraw.sty} around that are not distinguished by a
version number. Some of them present problems when used with JaxoDraw, please
try to use the one that we distribute with our program before reporting problems
regarding {\tt axodraw}.
Since we do not maintain a mailing list, we will try to make all messages of
general interest available on our Web-site 

\verb+http://altair.ific.uv.es/~JaxoDraw/home.html+

\noindent
Please check these pages and also the FAQ and Known problems sections of this
document before reporting any bugs.


\subsection{License}
\label{License}

\begin{verbatim}

Copyright (C) 2003 Daniele Binosi and Lukas Theussl
 
JaxoDraw is free software; you can redistribute it and/or 
modify it under the terms of the GNU General Public License 
as published by the Free Software Foundation; either version 2 
of the License, or (at your option) any later version.
 
JaxoDraw is distributed in the hope that it will be useful, 
but WITHOUT ANY WARRANTY; without even the implied warranty of
MERCHANTABILITY or FITNESS FOR A PARTICULAR PURPOSE.
See the GNU General Public License for more details.
 
A copy of the GNU General Public License can be found in the 
file GNU-LICENSE that is distributed along with this program.


JaxoDraw includes a copy of J. Vermaseren's axodraw.sty file
(with the kind permission of the author).
The file FORM-LICENSE in the distribution home directory reproduces
the author's license agreement as published at
\end{verbatim}

    \verb+http://www.nikhef.nl/~form/license.html+

\begin{verbatim}
The .jre version of JaxoDraw includes a copy of the Java Runtime Environment.
This is redistributed under the terms of conditions as published at
\end{verbatim}

    \verb+http://java.sun.com/j2se/1.4.1/j2re-1_4_1_03-license.html+

\begin{verbatim}
This product includes code licensed from RSA Security, Inc.
Some portions licensed from IBM are available at
\end{verbatim}

    \verb+http://oss.software.ibm.com/icu4j/+

\begin{verbatim}
PostScript is a trademark of Adobe Systems Incorporated.

\end{verbatim}


\section{Using JaxoDraw}
\label{usage}

\subsection{Terminology}
\label{Terminology}

\begin{itemize}
\item[] {\bf Object} \\
          An object is the collection of points, with optional associated 
          values, that makes up one entity of a Feynman diagram. 
	      Examples are blobs, lines, arcs, boxes, loops, etc. 
	      The associated values can be used to change the appearance of an 
	      object, like color, line width, photon amplitude, 
          and several other features. 
	      
\item[]  {\bf Graph} \\
          A graph is the collection of objects (lines, arcs, ...), 
          together with titles, comments, layout options, etc. drawn to 
	      display the Feynman diagram. 

\item[] {\bf Parameters} \\
          Parameters are the settings of symbols, line styles, colors, 
          fonts, etc. used to define graphs and the display of the 
	      active objects.

\item[] {\bf Handles} \\
          When the program goes into Edit mode (any mode that allows the
          modification of any parameters of any object) little red squares are
          displayed on certain points of every object (for instance on the end
          points of lines). When the user clicks on
          one of these handles, the corresponding edit operation is being
          carried out on the chosen object.
\end{itemize}


\subsection{Execution}
\label{Execution}

The most convenient way to start JaxoDraw depends on your operating system and
on how you installed the program. See the Installation 
section~\ref{Installation} for generic instructions to run JaxoDraw. 

In general, if you compiled the package yourself from sources and put all the 
binary {\tt .class} files (together with the {\tt doc/} and {\tt icons/} 
sub-folders) in a sub-directory called {\tt JaxoDraw/}, you can
start JaxoDraw by the command

{\tt java JaxoDraw/JaxoDraw}

\noindent
in the current directory. Supposing you have Java installed on your system, 
this will work on any platform.


\subsection{Command line parameters}
\label{Commandlineparams}

The current version supports the following command line parameters:

\begin{itemize}
\item[] {\bf \verb~--~version} \\
        Prints out the version number of JaxoDraw.
\item[] {\bf \verb~--~help} \\
        Prints out some usage info on the standard output.
\item[] {\bf \verb~--~info}\\
        Prints out some information about your system. 
\item[] {\bf \verb~-~verbose} \\
        Turns on verbose error messaging (default in the current version).
\item[] {\bf \verb~-~quiet}\\
        Turns off verbose error messaging.
\end{itemize}

\noindent
By default, all parameters starting with {\bf \verb~--~} do not pop up the 
graphical user interface of JaxoDraw.

Furthermore, if you have saved an XML file with a JaxoGraph in an earlier
session, you may read in this graph directly on the command line by supplying
the file name as an argument (the extension of the file has to be {\tt .xml}).


\subsection{Drawing}
\label{Drawing}

Drawing Feynman diagrams with JaxoDraw is pretty easy and self-explaining. The
program has been designed with the main strategy to be easy to use. In
particular, if you are familiar with the 
\htmladdnormallinkfoot{xfig}{http://www.xfig.org/} program, you will have little
problems to get used to JaxoDraw (even though there are a few differences, in
particular when drawing arcs, see below). In general, to draw an element of a 
Feynman 
diagram, you first choose the drawing mode by clicking on the corresponding
button in the button panel, and then draw the object by left mouse-clicking and
dragging on the canvas. Drawn objects may then be moved/resized or edited by
choosing the corresponding button in the edit button panel and then clicking on
one of the handles specifying the object. 

A few things to note:

\begin{itemize}
\item[] Arcs are always drawn first with a default angle of 
        180\mbox{$^{\circ}$} (a different default angle may be chosen in the
        Preferences dialog). The angle can then only be changed via the 
        editing menu.
\item[] Any operation that changes any attribute of an object (move, resize,
        edit, ...) will automatically put the object in the foreground. 
\item[] It is a good idea to use the refresh button from time to time,
        especially if there are a lot of objects on the screen and if you are
        using antialising.
\end{itemize}


\subsection{Setting resources}
\label{Settingresources}

JaxoDraw allows the permanent setting of preferences via the Preferences menu. 
If you press the "Save" button for the first time in the Preferences dialog, a
corresponding preferences file called {\tt .Jaxorc} will be created in your 
home directory. This file is read automatically every time JaxoDraw is started.
It is in an XML format that may be edited manually if you know what you are
doing, the preferred way of editing it is via the above mentioned  Preferences 
menu dialog of the graphical user interface. See the Preferences menu item of 
section~\ref{menubar} for more
information on the items that may be saved on a permanent basis.


\subsection{Colors}
\label{colors}

\noindent
In the current version of JaxoDraw, the user may choose from a set of 84 colors
that are presented in a convenient color chooser panel if the user clicks an
object in color mode. The colors include all the 68 colors defined by the
{\tt colordvi} \LaTeX~class (on a standard TeTeX distribution, these may be 
found in {\tt /usr/share/texmf/tex/plain/dvips/colordvi.tex}) 
and 16 gray scales.  If you produce figures with color via the 
{\tt latex \verb|->| dvips} commands of JaxoDraw, these colors will be used as 
defined in the {\tt colordvi}
style file. For direct postscript output, we have tried to reproduce as closely
as possible the RGB values of these colors, but since there are no complete RGB
specifications (for free), the output will not be exactly the same as in the 
\LaTeX~case.

As a reference, we include two files in the source distribution of JaxoDraw,
that illustrate the differences. The {\tt latexcolor.ps} file
in the {\tt JaxoDraw/doc/} directory gives a collection of all the colors 
present in {\tt colordvi} as produced by {\tt latex \verb|->| dvips}. 
The file {\tt pscolor.ps} in the same directory gives the corresponding 
collection as produced by direct postscript output.


\subsection{Text}
\label{text}

There are two ways of entering text in JaxoDraw: Postscript text mode and 
\LaTeX~text mode. Even though they may be used at the same time in a graph,
they will appear mutually exclusive in any derived output.

\begin{itemize}
\item[] {\bf Postscript text mode} \\
          When entering the postscript text mode, the user may enter a text string
          that will appear directly on the screen and in any direct postscript
          output (i.e., also in any printer output). It will not appear in any
          output created via {\tt latex \verb|->| dvips}. In edit mode, the user may choose
          the text size and font of the text object. 
          A set of Greek characters is available via a syntax that is derived
          from the corresponding \LaTeX~commands:
          
          \begin{center}
          \begin{tabular}{clclclcl}
          $\alpha$   & \verb|\alpha|   & $\lambda$   & \verb|\lambda|  & $\upsilon$  & \verb|\upsilon| & $\Lambda$ & \verb|\Lambda| \\
          $\beta$    & \verb|\beta|    & $\mu$       & \verb|\mu|      & $\phi$      & \verb|\phi|     & $\Xi$     & \verb|\Xi|     \\
          $\gamma$   & \verb|\gamma|   & $\nu$       & \verb|\nu|      & $\chi$      & \verb|\chi|     & $\Pi$     & \verb|\Pi|     \\
          $\delta$   & \verb|\delta|   & $\xi$       & \verb|\xi|      & $\psi$      & \verb|\psi|     & $\Sigma$  & \verb|\Sigma|  \\
          $\epsilon$ & \verb|\epsilon| & $o$         & \verb|o|        & $\omega$    & \verb|\omega|   & $\Phi$    & \verb|\Phi|    \\
          $\zeta$    & \verb|\zeta|    & $\pi$       & \verb|\pi|      & $\vartheta$ & \verb|\vartheta|& $\Psi$    & \verb|\Psi|    \\
          $\eta$     & \verb|\eta|     & $\rho$      & \verb|\rho|     & $\varphi$   & \verb|\varphi|  & $\Omega$  & \verb|\Omega|  \\
          $\theta$   & \verb|\theta|   & $\varsigma$ & \verb|\varsigma|& $\Gamma$    & \verb|\Gamma|   &           &                \\
          $\iota$    & \verb|\iota|    & $\sigma$    & \verb|\sigma|   & $\Delta$    & \verb|\Delta|   &           &                \\
          $\kappa$   & \verb|\kappa|   & $\tau$      & \verb|\tau|     & $\Theta$    & \verb|\Theta|   &           &   
          \end{tabular} 
          \end{center}
	      
          Note that no \$ signs are necessary for these commands (any \$ signs
          will appear verbatim on the screen). If the user
          enters  a string starting with a "\verb|\|" that is not recognized as
          a valid Greek letter, it will be replaced by a question mark "?". 
          In the current version of JaxoDraw it is not possible to do super- or
          subscripts, this will be implemented in some future version (see the
          wish list in section~\ref{Wishlist}).
 
\item[]  {\bf \LaTeX~text mode} \\
          When entering the \LaTeX~text mode, the user may enter a text string
          that will appear only in the \LaTeX~output file and any files created
          from it via {\tt latex \verb|->| dvips}. Like that all the commands 
          known to \LaTeX~in math mode are available to the user. 
          The position of the text will be marked on the
          screen by an icon that identifies it as a \LaTeX~text object. This
          icon does not appear in direct postscript or printing output. 
          Note that the \LaTeX~text string will automatically be put between \$
          signs, so the text will always be in \LaTeX~math mode. If you want a
          normal font in \LaTeX~text mode, you should use \verb|{\rm }|.
          Note also that your input here is the only possible source of errors
          in your \LaTeX~source code. If you get any \LaTeX~compilation errors,
          check your \LaTeX~text objects first.
          In edit mode, the user may choose the \LaTeX~font size and the
          alignment with respect to the current position of the text object. 

\end{itemize}


\section{Screen elements of JaxoDraw}
\label{elements}
The screen of JaxoDraw is divided into five main sections:

\begin{itemize}
\item[] The menu bar on top
\item[] The tool bar just below the menu bar
\item[] The button panel on the left
\item[] The status bar on bottom
\item[] The drawing area (the canvas) in the center
\end{itemize}

\noindent
The tool and the status bar may optionally be switched off in the preferences
dialog. In the following we will describe each of the above sections in greater 
detail.

\subsection{The menu bar}
\label{menubar}

The menu bar contains four main menu bar items: File, Edit, Options and Help. 

\paragraph{File}

\begin{itemize}
\item[] {\bf New} \\
Starts a new JaxoDraw graph, abandoning the current plot. 
\item[] {\bf Open} \\
Open an existing JaxoDraw file, abandoning the current plot. This pops up a 
file chooser dialog where the user may indicate an XML file that was stored in
an earlier session.
\item[] {\bf Save} \\
Save the current plot using the last specified name. If no name is specified, a
file chooser menu is popped up. The current graph is then saved in an XML format
that may be opened in a later session.
\item[] {\bf Save As}\\ 
The same as Save, but always
pops up a file chooser menu to save the current plot under the chosen name.
\item[] {\bf Describe}\\ 
Add a text description to a graph. This will appear as a comment in all output
files.
\item[] {\bf Export} \\ 
Pops up a dialog where the user may choose among several export file formats. 
These ones include: LaTeX \verb|->| EPS (to produce an encapsulated postscript 
({\tt eps}) file via {\tt latex \verb|->| dvips}), LaTeX (to produce a text file
containing \LaTeX~source code), Postscript Portrait, Postscript Landscape and
EPS (to produce direct postscript output via Java's internal postscript
interface in portrait, landscape and {\tt eps} format, respectively). 
Pressing the Export button will pop up a file chooser dialog to enter a 
file name for the chosen export format.

In addition, there is a button that allows to preview any of the above output
formats. Note that in order to preview any of the postscript exports, you will
need to indicate a postscript previewer in the preferences dialog (since there
is no Java internal postscript renderer). For previewing output in text format,
you may still indicate a preferred text editor but if you do not do so, a Java
internal text previewer is used by default.
\item[] {\bf Print} \\
Prints the current graph to a specified printer or postscript file. This opens
the standard Java printer dialog where any installed and configured printers are
detected automatically. Note that printing to a file should be equivalent to the
corresponding Export - Postscript option.
\item[] {\bf Quit} \\
Exits JaxoDraw. 

\end{itemize}

\paragraph{Edit}

\begin{itemize}
\item[]  {\bf Undo} \\
Cancels the last operation. Note that multiple Undo's are not possible.
\item[]  {\bf Clear} \\
Clears this graph. This only removes the visible objects from the screen, it
does not affect any values associated with the graph.
\item[]  {\bf Move} \\
Goes into move mode. Displays handles where the user may grab an object and move
it by dragging it over the screen.
\item[]  {\bf Resize} \\
Goes into Resize mode. Displays handles where the user may grab an object and 
resize it by dragging a specified point. Note that loops and arcs cannot be
resized from the center point.
\item[]  {\bf Copy} \\
Goes into Copy mode. Displays handles where the user may grab an object that
will be duplicated by an exact copy that the user may drag to a different
location.
\item[]  {\bf Color} \\
Goes into Color mode. Displays handles where the user may grab an object which
pops up a dialog to change the color of the object. See section~\ref{colors} for
more information on colors.
\item[]  {\bf Edit} \\
Goes into Edit mode. Displays handles where the user may grab an object which
pops up a dialog to change the parameters of the object. Which parameters may be
edited depends on the object.
\item[]  {\bf Delete} \\
Goes into Delete mode. Displays handles where the user may grab an object to
delete it.
\item[]  {\bf Background} \\
Goes into Background mode. Displays handles where the user may grab an object to
put it into the background.
\item[]  {\bf Foreground} \\
Goes into Foreground mode. Displays handles where the user may grab an object to
put it into the foreground.
\end{itemize}

\paragraph{Options}

\begin{itemize}
\item[] {\bf Look and Feel} \\
Lets the user choose a Look and Feel for the current session. Note that some
LAFs may not be available on your system and that there might be differences in
some layouts, in particular with icons in the tool bar.
\item[] {\bf Vertex types} \\
Lets the user choose the type of vertex to be drawn when in Vertex mode. By
default, it is a black dot, other vertices currently supported are a circle with
a cross, a square and a cross. Choosing one vertex type will change the icon of
the Vertex button in the button panel to the corresponding vertex. 
\item[] {\bf Show Toolbar} \\
Lets the user choose whether the toolbar is visible or not.
\item[] {\bf Show Statusbar} \\
Lets the user choose whether the statusbar is visible or not.
\item[] {\bf Antialias on} \\
Lets the user choose whether to use antialiasing or not. The graphics quality is
usually better with antialiasing turned on. This goes with the cost that
graphics rendering may be slower on some machines and you may need to refresh
the screen from time to time, especially after a number of editing operations.
\item[] {\bf Arrow} \\
Lets the user choose whether arrows should be drawn on all objects 
that support them.
\item[] {\bf Preferences} \\
Pops up a dialog where the user may choose several settings to be saved on a
permanent basis. The first group of settings are the default HTML viewer (to
view the JaxoDraw documentation in HTML format), a default text editor (used for
previewing \LaTeX~text output) and a default postscript viewer (used for
previewing the printer or direct postscript output). Note that you need
to specify a default postscript viewer in order to view postscript files from
within JaxoDraw because Java does not have an internal possibility to render
postscript files. Contrary to that, if you do not specify any default HTML 
viewer or text editor, previews will still be possible with the Java internal
HTML and text rendering mechanisms which will be used by default. 

You may then choose the default Look and Feel, the default size of the grid, the
default line width, the default opening angle for arcs, as well as the initial
screen size. Finally you may determine whether the tool- and the status bar are
visible by default, whether antialiasing should be used by default and whether 
arrows should be drawn by default on all objects that support them.

Clicking OK will apply the specified values for the current session without
saving them in the system configuration file {\tt .Jaxorc}, clicking Save will
save the settings without applying them to the current session, the button 
Clear only clears the text fields of the default previewers, Reset restores all
the values to their current default settings and Cancel closes the Preferences
dialog without applying any changes.

See section~\ref{Settingresources} for more information on setting resources.
\end{itemize}

\paragraph{Help}
\begin{itemize}
\item[] {\bf About} \\
Gives some information about the version of JaxoDraw you are using.
\item[] {\bf User guide} \\
Pops up a new window with this user guide in HTML format. If a default HTML
viewer has been chosen in the Preferences dialog, it will be used, otherwise a
Java internal previewer is used by default.
\item[] {\bf System info} \\
Gives some information about your system (current user, operating system, Java
installation).
\end{itemize}


\subsection{The tool bar}
\label{toolbar}

The tool bar may be switched on and off in the Options menu item. If it is
switched on, the tool bar contains icons whose action is identical to the 
corresponding menu entries: New, Open, Save, Save As, Describe, Export, Print. 
There is furthermore one icon that does a Latex \verb|->| EPS preview, 
and on the right there is an icon to pop up the user guide.


\subsection{The button panel}
\label{buttonpanel}

The button panel on the left of the screen is divided into the following
subsections:

\paragraph{Particle buttons}

There is one button for each particle type: fermion (straight line), scalar
(dashed line), ghost (dotted line), photon (wiggled line) and gluon (pig-tailed
line); and the three object types: lines, arcs and loops. When one of these
buttons is clicked the program goes into the corresponding drawing mode, i.e.,
no handles are shown on the screen and the user may click on the canvas to start
drawing the corresponding object.

\paragraph{Miscellaneous buttons}

There are buttons for drawing blobs (ellipses), boxes, vertices and zig-zag
lines, as well as buttons that allow the insertion of postscript text and 
\LaTeX~text into the graph. See section~\ref{text} above for information on 
postscript- and \LaTeX~text mode. 
When one of these buttons is clicked the program goes into the corresponding 
drawing mode.

\paragraph{Action buttons}

These are the buttons that lead to an immediate action: Undo and Clear have the
same effect as the corresponding entries in the Edit menu, while the Refresh
button leads to a redrawing of the screen. This is especially useful if
antialising is used.

\paragraph{Edit buttons}
There are a number of buttons (Move, Resize,
Copy, Color, Edit, Delete, Background and Foreground) whose actions are equivalent to
the ones described in the Edit menu panel section. When one of these
buttons is clicked the program goes into the corresponding edit mode.

\paragraph{Grid and exit button}
The grid button turns on the grid so that the user can choose only certain points
for placing his objects. Note that this will not change any objects already
present on the screen. The exit button quits JaxoDraw.


\subsection{The status bar}
\label{statusbar}

The status bar may be switched on and off in the Options menu item. If it is
switched on, the status bar contains three areas: one to display the current
file (if any), one to display the current drawing mode and one that displays the
current coordinates of the cursor on the canvas.


\subsection{The canvas}
\label{canvas}

This is the main drawing area. After choosing a drawing mode from the button
panel, the user may draw the corresponding object by left-clicking and dragging
on the canvas.


\section{Known problems and limitations}
\label{problems}

This section gives a list of bugs and limitations that were known at the time of
first publication of JaxoDraw-1.0. Please check the Bugs section of our 
\htmladdnormallinkfoot{Web page}
{http://altair.ific.uv.es/\~{}JaxoDraw/bugs.html}
for an updated version of this document. Note that not all points are
necessarily real bugs, we regard this just as a collection of features that do 
not work exactly the way we would like to.

\subsection{Bugs}
\label{Bugs}

\begin{itemize}

\item[*]  If a \LaTeX~text file is previewed with the Java internal previewer 
          (i.e. without having set a custom previewer in the preferences), the 
          page displays but cannot be scrolled. When the export chooser menu is 
          closed, it becomes scrollable (this may be used as a workaround).

\item[*]  Choosing a different Look and Feel in the Preferences dialog of
          the Options menu item and pressing the OK button, does not 
          update the Look and Feel of the current session. Workaround: use the
          Look and Feel dialog of the Options menu instead.

\item[*]  When using IBM's Runtime Environment, the program may be executed
          and works fine for most parts but presents some peculiarities: the
          layout of pop up windows is not always the same and XML output
          serializes the bounding boxes of objects that are explicitly marked as
          transient in the source code. This has been reported to us for version
          number {\tt 1.4.1} of IBM's SDK. For us, this appears to be an 
          incompatibility between SUN's and IBM's Runtime Environments. 
          On the other hand, the program compiles fine with IBM's jikes 
          compiler (tested with version {\tt 1.13}).

\item[*]  If your \LaTeX~compilations give errors complaining about unknown
          commands like \begin{ttfamily}\verb|\SetColor{}|\end{ttfamily},
          you will probably have to update your {\tt axodraw.sty} file to a
          more recent version (the one included in our 
          distribution will work).

\item[*]  There is a bug in {\tt axodraw} concerning gluon loops. In fact there
          is no command for drawing gluon loops, but if you draw a gluon arc
          with 360\mbox{$^{\circ}$}, then the gluon wiggles do not close
          correctly. Since this is a bug in {\tt axodraw}, there is no
          workaround except not to use gluon loops, or, if you need nicely
          closing loops, to use the direct postscript output of JaxoDraw.

\item[*]  Running an internal \LaTeX~compilation without {\tt axodraw}
          installed (or \LaTeX in the case you are a Windows user),  
          will hang the program. Please check appendix~\ref{axodraw} for
          information about installing {\tt axodraw}.

\end{itemize}

\subsection{Wish list}
\label{Wishlist}

The following are requested features that we will hopefully implement in a 
future version of JaxoDraw:

\begin{itemize}
\item[*] Enhance the TextParser class to allow sub- and superscripts in text objects
\item[*] Implement a triangle vertex
\item[*] Improve the drawing of arcs
\item[*] Let the user include custom packages in the \LaTeX~output
\item[*] Add the possibility to move/copy a group of objects
\item[*] Allow to work with several graphs at a time
\item[*] Show the \LaTeX~text in a pop-up window when rolled over
\end{itemize}


\section{Documentation}
\label{Docs}

This section gives a list of hints and tricks as well as a list of frequently
asked questions that were known at the time of
first publication of JaxoDraw-1.0. Please check the Docs section of our 
\htmladdnormallinkfoot{Web page}{http://altair.ific.uv.es/\~{}JaxoDraw/docs.html},
for an updated version of this document.

\subsection{Tips and tricks}
\label{tips}

\begin{itemize}

\item[*] Note that postscript files produced by Export - EPS are considerably
         larger in size than the same files generated via 
         {\tt latex \verb|->| dvips}. This is
         due to the way how Java handles the postscript printing internally.
         Keep this in mind if you want to include a bunch of small figures in
         your document: it is then probably preferable to use the \LaTeX~output.

\item[*] It is a good idea to use the refresh button from time to time,
         especially if there are a lot of objects on the screen and if you are
         using antialising.

\item[*] Instead of producing EPS figures and including them into your document,
         you may as well cut-and-paste the \LaTeX~output of JaxoDraw into 
         your own \LaTeX~source code. Like that you may avoid the proliferation
         of numerous postscript files to be distributed with your source code.
         Note however that you will have to include {\tt axodraw} in the header of
         your \LaTeX~file 
         (put \begin{ttfamily}\verb|\usepackage{axodraw}|\end{ttfamily} 
         somewhere
         before \begin{ttfamily}\verb|\begin{document}|\end{ttfamily}) and 
         you will probably have to
         distribute the {\tt axodraw.sty} file along with your source code
         because it is not part of any standard \LaTeX~distribution.

\item[*] To add multiple arrows to objects (such as loops, for example), 
         or to draw arrows on objects that do not support them 
         (photon and gluon objects in particular),
         use fermion lines with very small length (5 points should do the job). 
         Use the edit menu, to give the arrow the inclination you need and 
         then move it to the wanted location.

\end{itemize}


\subsection{FAQ}
\label{faq}

No questions so far ...


\section{History}
\label{history}

\begin{tabular}{rl}
1.9.2003 & Released JaxoDraw-1.0
\end{tabular}


\section{Credits}
\label{credits}

We are grateful to Prof. Arcadi Santamaria for numerous helpful remarks and
moral support during the development of JaxoDraw.
We also acknowledge Prof. Jos Vermaseren for his kind permission to use and
distribute his axodraw style file along with JaxoDraw.\\

\noindent{Work partially supported by the grant FPA-2002-00612.}


\begin{appendix}

\section{Installing axodraw.sty}
\label{axodraw}

{\bf Note}: It is not necessary to install {\tt axodraw} in order to run
JaxoDraw. You will just not be able to use the LaTeX/LaTeX - EPS export options 
but you may still generate direct postscript output of your Feynman diagrams.
Beware however that in the current version, trying to run an internal
\LaTeX~compilation without {\tt axodraw} installed, will hang the program. 
See the Bugs section~\ref{Bugs}.

\medskip
\noindent 
In the current version of JaxoDraw we distribute a copy of J. Vermaseren's 
{\tt axodraw} package (with kind permission of the author) in the
distribution home directory. You have to 
install {\tt axodraw.sty} such that \LaTeX~can find 
it on your system. This appendix describes how to do that.

\medskip
\noindent 
First get J. Vermaseren's {\tt axodraw} package from  

\hturl{http://www.nikhef.nl/~form/FORMdistribution/axodraw/}

\noindent 
or from the JaxoDraw distribution home directory.
Please refer to the {\tt axodraw} user guide for a detailed documentation of the 
package. We shall only outline here how you make {\tt axodraw} available on your
system and how you use it with the \LaTeX~output from JaxoDraw.

\subsection{Linux instructions}

For installation, you have two options: if you intend to use {\tt axodraw} just for
yourself on a multi-user platform, you may install it locally; if you want to
make it available for all the users on the system, you should do a global
installation. Note that you will need root privileges for a global installation.

\medskip
{\bf Installing {\tt axodraw} locally}

\noindent 
The easiest way to use {\tt axodraw} is to put the {\tt axodraw.sty} file in the same
directory as your \LaTeX~source file (like the one produced by JaxoDraw via the
Export \verb|->| LaTeX command). This is usually the same directory where you
execute the program (but note that you cannot execute JaxoDraw from a different
directory in this case). You can then run {\tt latex} on your source file as 
usual, the style file will be found because the current directory is by 
default in the {\tt TEXINPUTS} search path.

An alternative (better) way is to put the style file in a special directory 
(this is particularly useful if you have several style files which are not 
part of your standard \LaTeX~distribution). Let's say you put it into the directory 
{\tt latex/} in your home directory. You then have to set the {\tt TEXINPUTS} 
variable to this path. If you are using bash, just do

{\tt export TEXINPUTS=\$HOME/latex//:}

\noindent 
(the //: at the end tells \LaTeX~also to look into
sub-directories of this path).
If you want to make that permanent, you should put this line into your 
{\tt .bashrc} file. You will then be able to start JaxoDraw from any directory,
independently of the location of {\tt axodraw.sty}.

\medskip
{\bf Installing {\tt axodraw} system-wide}

\noindent 
Installing {\tt axodraw} system wide is very easy. Just put the style file
somewhere in the global search path of your \LaTeX~distribution 
(on Redhat Linux
typically {\tt /usr/share/texmf/tex/latex/misc/}) and update the \TeX~database
with

{\tt mktexlsr}

\noindent 
(you will have to be root for doing this).

\subsection{Windows instructions}

Under Windows, you have to do basically the same as under Linux.  First put your
{\tt axodraw.sty} file into the MikTeX search tree (somewhere under 
{\tt /texmf/tex/latex/}, replace the slash by a backslash!) and
update your database with the command {\tt mktexlsr} in the 
{\tt /texmf/tex/miktex/bin} directory.

\end{appendix}



\end{document}